\documentclass[12pt,preprint]{aastex}
\renewcommand\bv{\mbox{\boldmath $v$}}
\renewcommand\>{{\rangle}}
\newcommand\qe{{\tilde{q}}}

\newcommand\bnabla{\mbox{\boldmath $\nabla$}}

\newcommand\ex{\mbox{\boldmath $\hat{x}$}}
\newcommand\ey{\mbox{\boldmath $\hat{y}$}}

\newcommand\bO{\mbox{\boldmath $\Omega$}}

\newcommand\<{{\langle}}

\newcommand{\dv}[2]{\frac{d#1}{d#2}}
\newcommand{\be}{\begin{equation}}
\newcommand{\ee}{\end{equation}}
\shortauthors{Johnson \& Gammie}
\shorttitle{Nonlinear Radial Stability}
\begin{document}

\title{Nonlinear Stability of Thin, Radially-Stratified Disks}

\author{Bryan M. Johnson and Charles F. Gammie}

\affil{Center for Theoretical Astrophysics,
University of Illinois at Urbana-Champaign,
1110 West Green St., Urbana, IL 61801}

\begin{abstract}

We perform local numerical experiments to investigate the nonlinear
stability of thin, radially-stratified disks.  We demonstrate the
presence of radial convective instability when the disk is nearly in
uniform rotation, and show that the net angular momentum transport
is slightly inwards, consistent with previous investigations of
vertical convection.  We then show that a convectively-unstable
equilibrium is stabilized by differential rotation.  Convective 
instability is determined by the Richardson number 
${\rm Ri} \equiv N_r^2/(\qe \Omega)^2$, where $N_r$ is the radial 
Brunt-V$\ddot{\rm{a}}$is$\ddot{\rm{a}}$l$\ddot{\rm{a}}$ frequency 
and $\qe \Omega$ is the shear rate.  Classical convective instability in 
a nonshearing medium (${\rm Ri} \rightarrow -\infty$) is suppressed 
when ${\rm Ri} \gtrsim -1$, i.e. when the shear rate becomes greater 
than the growth rate.  Disks with a nearly-Keplerian rotation profile 
and radial gradients on the order of the disk radius have ${\rm Ri} 
\gtrsim -0.01$ and are therefore stable to local nonaxisymmetric 
disturbances.  One implication of our results is that the ``baroclinic'' 
instability recently claimed by Klahr \& Bodenheimer is either global 
or nonexistent.  We estimate that our simulations would detect any 
genuine growth rate $\gtrsim 0.0025 \Omega$.

\end{abstract}

\keywords{accretion, accretion disks, solar system: formation, galaxies:
nuclei}

\section{Introduction}

In order for astrophysical disks to accrete, angular momentum must be
removed from the disk material and transported outwards.  In many disks,
this outward angular momentum transport is likely mediated internally by
magnetohydrodynamic (MHD) turbulence driven by the magnetorotational
instability (MRI; see \citealt{bh98}).  A key feature of this transport
mechanism is that it arises from a local shear instability and is therefore
very robust.  In addition,
MHD turbulence transports angular momentum {\it outwards}; some other forms
of turbulence, such as convective turbulence, appear to transport angular
momentum inwards \citep{sb96,cab96}.  The mechanism is only effective, however, if
the plasma in the disk is sufficiently ionized to be well-coupled to the
magnetic field \citep{bb94,hgb95,gam96,jin96,war99,fsh00,bt01,ss02a,ss02b,
sw03,sal04,kb04,des04}.  In portions of disks around young,
low-mass stars, in cataclysmic-variable disks in quiescence, and in X-ray
transients in quiescence \citep{sgbh00,gm98,men00}, the plasma may be too
neutral for the MRI to operate.  This presents some difficulties for
understanding the evolution of these systems, since no robust transport mechanism
akin to MRI-induced turbulence has been established for purely-hydrodynamic
Keplerian shear flows.

Such a mechanism has been claimed recently by \cite{kb03},
who find vortices and an outward transport of angular momentum in the
nonlinear outcome of their global simulations.  The claim is that this
nonlinear outcome is due to a {\it local} instability (the ``Global
Baroclinic Instability'') resulting from the presence of an equilibrium
entropy gradient in the radial direction.\footnote{The term ``global'' is 
employed by \cite{kb03} to highlight the fact that their equilibrium 
gradients do not contain any localized features; it does not refer to the 
nature of the instability.}  The instability mechanism
invoked \citep{klr04} is an interplay between transient amplification of
linear disturbances and nonlinear effects.  The existence of such a
mechanism would have profound implications for understanding the
evolution of weakly-ionized disks.

In a companion paper (\citealt{jg05}, hereafter Paper I), we have
performed a linear stability analysis for local nonaxisymmetric
disturbances in the shearing-wave formalism.  While the linear theory
uncovers no exponentially-growing instability (except for convective
instability in the absence of shear), interpretation of the results
is somewhat difficult due to the nonnormal nature of the linear
differential operators:\footnote{A nonnormal operator is one that is
not self-adjoint, i.e. it does not have orthogonal eigenfunctions.}
one has a coupled set of differential
equations in time rather than a dispersion relation, which results
in a nontrivial time dependence for the perturbation amplitudes
$\delta(t)$.  In addition, transient amplification does occur for
a subset of initial perturbations, and linear theory cannot tell us
what effect this will have on the nonlinear outcome.  For these
reasons, and in order to test for the presence of local nonlinear
instabilities, we here supplement our linear analysis with local
numerical experiments.

We begin in \S2 by outlining the basic equations for a local model of
a thin disk. In \S3 we summarize the linear theory results from Paper I.
We describe our numerical model and nonlinear results in \S\S4 and 5, and
discuss the implications of our findings in \S6.

\section{Basic Equations}

The most relevant results from \cite{kb03} are those that came out of their 
two-dimensional (radial-azimuthal) simulations, since the salient feature 
supposedly giving rise to the instability is a radial entropy gradient.  
The simplest model to use
for a local verification of their global results is the two-dimensional
shearing sheet (see, e.g., \citealt{gt78}).  This local approximation
is made by expanding the equations of motion in the ratio of the disk
scale height $H$ to the local radius $R$, and is therefore only valid
for thin disks ($H/R \ll 1$).  The vertical structure is removed by
using vertically-integrated quantities for the fluid variables.\footnote{This
vertical integration is not rigorous.}  The basic equations that one obtains are
\be\label{EQ1}
\dv{\Sigma}{t} + \Sigma \bnabla \cdot \bv = 0,
\ee
\be\label{EQ2}
\dv{\bv}{t} + {\bnabla P\over{\Sigma}} + 2\bO\times\bv - 2q\Omega^2 x \ex = 0,
\ee
\be\label{EQ3}
\dv{\,{\rm{ln}} S}{t} = 0,
\ee
where $\Sigma$ and $P$ are the two-dimensional density and pressure, $S
\equiv P \Sigma^{-\gamma}$ is the fluid entropy,\footnote{For a
non-self-gravitating disk the two-dimensional adiabatic index $\gamma =
(3\gamma_{3D} - 1)/(\gamma_{3D} + 1)$ (e.g. \citealt{ggn86}).} $\bv$ is
the fluid velocity and $d/dt$ is the Lagrangian derivative. The third and
fourth terms in equation (\ref{EQ2}) represent the Coriolis and centrifugal
forces in the local model expansion, where $\Omega$ is the local rotation
frequency, $x$ is the radial Cartesian coordinate and $q$ is the shear
parameter (equal to $1.5$ for a disk with a Keplerian rotation profile).
The gravitational potential of the central object is included as part of
the centrifugal force term in the local-model expansion, and we ignore the
self-gravity of the disk.

\section{Summary of Linear Theory Results}

An equilibrium solution to equations (\ref{EQ1}) through (\ref{EQ3}) is
\be\label{P0}
P = P_0(x),
\ee
\be
\Sigma = \Sigma_0(x),
\ee
\be\label{V0}
\bv \equiv \bv_0 = \left(-q\Omega x + \frac{P_0^\prime}{2\Omega
\Sigma_0}\right)\ey,
\ee
where a prime denotes an $x$ derivative.
One can regard the background flow as providing an effective shear rate
\be
\qe \Omega \equiv -v_0^\prime
\ee
that varies with $x$, in which case $\bv_0 = -\int^x \qe(s) ds \, \Omega
\ey$.  Due to this background shear, localized disturbances can be
decomposed in terms of ``shwaves'', Fourier modes in a frame comoving with
the shear.  These have a time-dependent radial wavenumber given by
\be\label{KXEFF}
\tilde{k}_x(t,x) \equiv k_{x0} + \qe(x)\Omega k_y t.
\ee
where $k_{x0}$ and $k_y$ are constants.  Here $k_y$ is the azimuthal
wave number of the shwave.

In the limit of zero stratification,
\be
P_0(x) \rightarrow constant,
\ee
\be
\Sigma_0(x) \rightarrow constant,
\ee
\be\label{V0U}
\bv_0 \rightarrow -q\Omega x \ey,
\ee
\be
\qe \rightarrow q,
\ee
and
\be\label{KX}
\tilde{k}_x \rightarrow k_x \equiv k_{x0} + q\Omega k_y t.
\ee
In Paper I, we analyze the time dependence of the shwave amplitudes for both
an unstratified equilibrium and a radially-stratified equilibrium.  As
discussed in more detail in Paper I, applying the shwave formalism to a
radially-stratified shearing sheet effectively uses a short-wavelength
WKB approximation, and is therefore only valid in the limit $k_y L \gg 1$,
where the background varies on a scale $L \sim H \ll R$.  The disk scale
height $H \equiv c_s \Omega$, where $c_s = \sqrt{\gamma P_0/\Sigma_0}$.

There are three nontrivial shwave solutions in the unstratified shearing
sheet, two nonvortical and one vortical.  The radial stratification
gives rise to an additional vortical shwave.  In the limit of 
tightly-wound shwaves ($|k_x| \gg k_y$), the nonvortical and vortical
shwaves are compressive and incompressive, respectively.  The former 
are the extension of acoustic modes to nonaxisymmetry, and to leading 
order in $(k_y L)^{-1}$ they are the same both with and without 
stratification.  Since the focus of our investigation is on convective 
instability and the generation of vorticity, we repeat here only the
solutions for the incompressive vortical shwaves and refer the
reader to Paper I for further details on the nonvortical shwaves.

In the unstratified shearing sheet, the solution for the incompressive shwave is given
by:
\be\label{IVX}
\delta v_x = \delta v_{x0}\frac{k_0^2}{k^2},
\ee
\be\label{IVY}
\delta v_y = -\frac{k_x}{k_y} \delta v_x
\ee
and
\be\label{IS}
\frac{\delta \Sigma}{\Sigma_0} = \frac{\delta P}{\gamma P_0} =
\frac{1}{i c_s k_y}\left(\frac{k_x}{k_y} \frac{\dot{\delta v}_x}{c_s}
+ 2(q - 1) \Omega \frac{\delta v_x}{c_s}\right),
\ee
where $k^2 = k_x^2 + k_y^2$, ($k_0,\delta v_{x0}$) are the values of 
($k, \delta v_x$) at $t=0$ and an overdot denotes a time 
derivative.\footnote{As
discussed in Paper I, this solution is valid for all time only in the short-wavelength limit ($k_y H \gg 1$); for $H k_y \lesssim O(1)$, an
initially-leading incompressive shwave will turn into a compressive
shwave near $k_x = 0$.}

The kinetic energy for a single incompressive shwave can be defined as
\be
E_k \equiv \frac{1}{2}\Sigma_0 (\delta v_x^2 + \delta v_y^2) =
\frac{1}{2}\Sigma_0 \delta v_{x0}^2 \frac{k_0^4}{k_y^2 k^2},
\ee
an expression which varies with time and peaks at $k_x = 0$.
If one defines an amplification factor for an individual shwave,
\be\label{AMP}
{\cal A } \equiv \frac{E_k(k_x = 0)}{E_k(t = 0)} =
1 + \frac{k_{x0}^2}{k_y^2},
\ee
it is apparent that an arbitrary amount of transient amplification in
the kinetic energy of an individual shwave can be obtained as one 
increases the amount of swing for a leading shwave ($k_{x0} \ll -k_y$).

This transient amplification of local nonaxisymmetric disturbances is 
reminiscent of the ``swing amplification'' mechanism that occurs in 
disks that are marginally-stable to the axisymmetric gravitational 
instability \citep{glb65,jt66,gt78}.  In that context, nonaxisymmetric 
shwaves experience a short period of exponential growth near $k_x = 0$ 
as they swing from leading to trailing.  In order for this mechanism 
to be effective in destabilizing a disk, however, a feedback mechanism 
is required to convert trailing shwaves into leading shwaves (see, e.g., 
\citealt{bt87} in the context of shearing waves in galactic disks).  The 
arbitrarily-large amplification implied by equation
(\ref{AMP}) has led some authors to argue for a bypass transition to
turbulence in hydrodynamic Keplerian shear flows \citep{cztl03,ur04,amn04}.
The reasoning is that nonlinear effects somehow provide the necessary
feedback.  We show in Paper I that an ensemble of incompressive shwaves drawn
from an isotropic, Gaussian random field has a kinetic energy that is a
constant, independent of time.  This indicates that a random set
of vortical perturbations will not extract energy from the mean shear.  It
is clear, however, that the validity of this mechanism as a transition to
turbulence can only be fully explored with a nonlinear analysis or numerical 
experiment.  To date, no such study has demonstrated a {\it transition} to
turbulence in Keplerian shear flows.

In the presence of radial stratification, there are two linearly-independent
incompressive shwaves.  The radial-velocity perturbation satisfies the
following equation (we use a subscript $s$ to distinguish the stratified
from the unstratified case):
\be\label{BOUSSVX2D}
(1 + \tilde{\tau}^2)\dv{^2\delta v_{xs}}{\tilde{\tau}^2} + 4 \tilde{\tau}
\dv{\delta v_{xs}}{\tilde{\tau}} + ({\rm{Ri}} + 2)\delta v_{xs} = 0,
\ee
where
\be
\tilde{\tau} \equiv \tilde{k}_x/k_y = \qe\Omega t + k_{x0}/ky
\ee
is the time variable and
\be\label{RICH}
{\rm Ri} \equiv \frac{N_x^2}{(\qe\Omega)^2}
\ee
is the Richardson number, a measure of the relative importance of buoyancy
and shear \citep{jwm61,how61,chi70}.\footnote{As discussed in Paper I,
equation (\ref{BOUSSVX2D}) is the same equation that one obtains for the
incompressive shwaves in a shearing, stratified atmosphere.}  Here
\be
N_x^2 \equiv -\frac{c_s^2}{L_S L_P}
\ee
is the square of the local Brunt-V$\ddot{\rm{a}}$is$\ddot{\rm{a}}$l$\ddot{\rm{a}}$
frequency in the radial direction, where $L_P \equiv \gamma P_0/P_0^\prime$
and $L_S \equiv \gamma S_0/S_0^\prime$ are the equilibrium pressure and 
entropy length scales in the radial direction.  The solutions for the other
perturbation variables are related to $\delta v_{xs}$ by
\be\label{XIY}
\delta v_{ys} = -\tilde{\tau} \delta v_{xs},
\ee
\be\label{SOX}
\frac{\delta \Sigma_s}{\Sigma_0} = \frac{1}{L_S} \int \delta v_{xs}\,dt
\ee
and
\be\label{SOLDH}
\frac{\delta P_s}{P_0} = \frac{\gamma \Omega}{i c_s k_y} \left[\qe
\tilde{\tau} \dv{}{\tilde{\tau}}\left(\frac{\delta v_{xs}}{c_s}\right) +
2(\qe - 1) \frac{\delta v_{xs}}{c_s}\right].
\ee

Since the solutions to equation (\ref{BOUSSVX2D}) are hypergeometric
functions, which have a power-law time dependence, it cannot in general be
accurately treated with a WKB analysis; there is no asymptotic region in
time where equation (\ref{BOUSSVX2D}) can be reduced to a dispersion
relation.  If, however, there  is a region of the disk where the effective
shear is zero, $\tilde{\tau} \rightarrow constant$ and equation
(\ref{BOUSSVX2D}) can be expressed as a WKB dispersion relation:
\be\label{DRQ0}
\omega^2 = \frac{k_y^2}{k_{x0}^2 + k_y^2}N_x^2,
\ee
with $\delta(t) \propto \exp(-i\omega t)$.
For $\qe \simeq 0$ and $N_x^2 < 0$, then, there is convective instability.
For disks with nearly-Keplerian rotation profiles and modest radial
gradients, $\qe \simeq 1.5$ and one would expect that the instability is
suppressed by the strong shear.  Due to the lack of a dispersion relation, 
however, there is no clear cutoff between exponential and oscillatory time 
dependence, and establishing a rigorous analytic stability criterion is difficult.

For $\qe \neq 0$, the asymptotic time dependence for each perturbation variable
at late times is
\be
\delta P_s \propto \delta v_{xs} \sim t^{\frac{\alpha - 3}{2}},
\ee
\be
\delta \Sigma_s \propto \delta v_{ys} \sim t^{\frac{\alpha - 1}{2}},
\ee
where
\be\label{ALPHA}
\alpha \equiv \sqrt{1 - 4 \, {\rm{Ri}}}.
\ee
The density and $y$-velocity perturbations therefore grow asymptotically for
$\alpha > 1$, i.e. ${\rm{Ri}} < 0$.  For small Richardson number, as is
expected for a nearly-Keplerian disk with modest radial gradients,
$\alpha \simeq 1 - 2{\rm Ri}$ and the asymptotic growth is extremely slow:
\be\label{ASYMP}
\delta \Sigma_s \sim \delta v_{ys} \sim t^{-{\rm Ri}}.
\ee
The energy of an ensemble of shwaves grows asymptotically as
$t^{2\alpha-1}$, or $t^{1-4{\rm Ri}}$ for small Ri.  The ensemble energy
growth is thus nearly linear in time for small Ri, independent of the sign
of Ri.\footnote{We show in Paper I that there is also linear growth in the
energy of an ensemble of compressive shwaves.}

The velocity perturbations are changed very little by a weak radial
gradient.  One would therefore expect that, at least in the linear
regime, transient amplification of the kinetic energy for an individual
shwave is relatively unaffected by the presence of stratification.
There is, however, an associated density perturbation in the stratified 
shearing sheet that is not present in the unstratifed sheet.\footnote{
The amplitude of the density perturbation in the unstratified sheet is 
an order-of-magnitude lower than the velocity perturbations in the 
short-wavelength limit; see equation(\ref{IS}).}  This results in 
transient amplification of the {\it potential} energy of an individual 
shwave.  We do not derive in Paper I a general closed-form expression 
for the energy of an ensemble of incompressive shwaves in the stratified 
shearing sheet, so it is not entirely clear what effect this 
qualitatively new piece of the energy will have on an ensemble of 
shwaves in the linear regime.

Due to the subtleties involved in interpreting the transient amplification 
of linear disturbances, and since the linear theory indicates weak 
power-law rather than exponential asymptotic growth in radially-stratified 
disks, we expect that stability will be decided by higher-order (nonlinear) 
interactions. This suggests a nonlinear, numerical study, which we describe 
here in the context of local numerical experiments in a radially-stratified 
shearing sheet.

\section{Numerical Model}

To investigate local nonlinear effects in a radially-stratified thin
disk, we integrate the governing equations (\ref{EQ1}) through (\ref{EQ3})
with a hydrodynamics code based on ZEUS \citep{sn92}.  This is a
time-explicit, operator-split, finite-difference method on a staggered
mesh.  It uses an artificial viscosity to capture shocks.  The
computational grid is $L_x \times L_y$ in physical size with $N_x \times
N_y$ grid cells, where $x$ is the radial coordinate and $y$ is the
azimuthal coordinate.  The boundary conditions are periodic in the
$y$-direction and shearing-periodic in the $x$-direction.  The shearing-box
boundary conditions are described in detail in \cite{hgb95}.  As described 
in \cite{mass00} and \cite{gam01}, advection by the linear shear flow can be
done by interpolation.  Rather than using a linear interpolation scheme as
in \cite{gam01}, we now do the shear transport with the same upwind 
advection algorithm used in the rest of the code.  This is less diffusive 
than linear interpolation, and the separation of the shear from the bulk 
fluid velocity means that one is not Courant-limited by large shear 
velocities at the edges of the computational domain.

We use the following equilibrium profile, which in general gives rise to 
an entropy that varies with radius:
\be
h_0(x) = h_a\left[1 - \epsilon \cos\left(\frac{2\pi x}{L_x}\right)\right] \;
\; , \; \; \Sigma_0(x) =
\left[\frac{h_0(\Gamma-1)}{\Gamma K}\right]^{\frac{1}{\Gamma-1}} \; \; , \;
\; P_0(x) = K \Sigma_0^\Gamma,
\ee
where $h_a$, $\epsilon$,
$\Gamma$ and $K$ are model parameters. The flow is maintained in
equilibrium by setting the initial velocity according to equation
(\ref{V0}).  This equilibrium yields a
Brunt-V$\ddot{\rm{a}}$is$\ddot{\rm{a}}$l$\ddot{\rm{a}}$ frequency
\be\label{N2}
N_x^2(x) = \frac{(\gamma-\Gamma) {h_0^\prime}^2}{\gamma (\Gamma-1) h_0},
\ee
which can be made positive, negative or zero by varying $\gamma-\Gamma$.

We fix some of the model parameters to yield an equilibrium profile that
is appropriate for a thin disk.  In particular, we want $H/L_P \sim H/R \ll
1$ in order to be consistent with our use of a razor-thin (two-dimensional)
disk model.  In addition, we want the equilibrium values for each fluid
variable to be of the same order to ensure the applicability of our linear
analysis.  These requirements can be met by choosing $K = 1$, $\epsilon =
0.1$, $L_x = 12$ and $h_a = \bar{c}_s^2 \Gamma/(\Gamma-1)$, where
$\bar{c}_s \equiv \sqrt{\< P_0/\Sigma_0 \>} \equiv 1$ is (to within a factor of
$\sqrt{\gamma}$) the $x$ average of the sound speed.  Since the equilibrium
profile changes with $\Gamma$, we choose a fixed value of $\Gamma = 4/3$,
which for $\gamma = \Gamma$ corresponds to a three-dimensional adiabatic
index of $7/5$.  These numbers yield $|H/L_P| \leq 0.2$.  Our unfixed model 
parameters are thus $L_y$, $q$ and $\gamma$.

The sinusoidal equilibrium profile we are using generates radial
oscillations in the shearing sheet, since the analytic equilibrium differs 
from the numerical equlibrium because of truncation error. We apply an
exponential-damping term to the governing equations in order to reduce the
spurious oscillations and therefore get cleaner growth-rate measurements.
We damp the oscillations until their amplitude is equal to that of
machine-level noise, and subsequently apply low-level random perturbations
to trigger any instabilities that may be present.

\begin{figure}
\plottwo{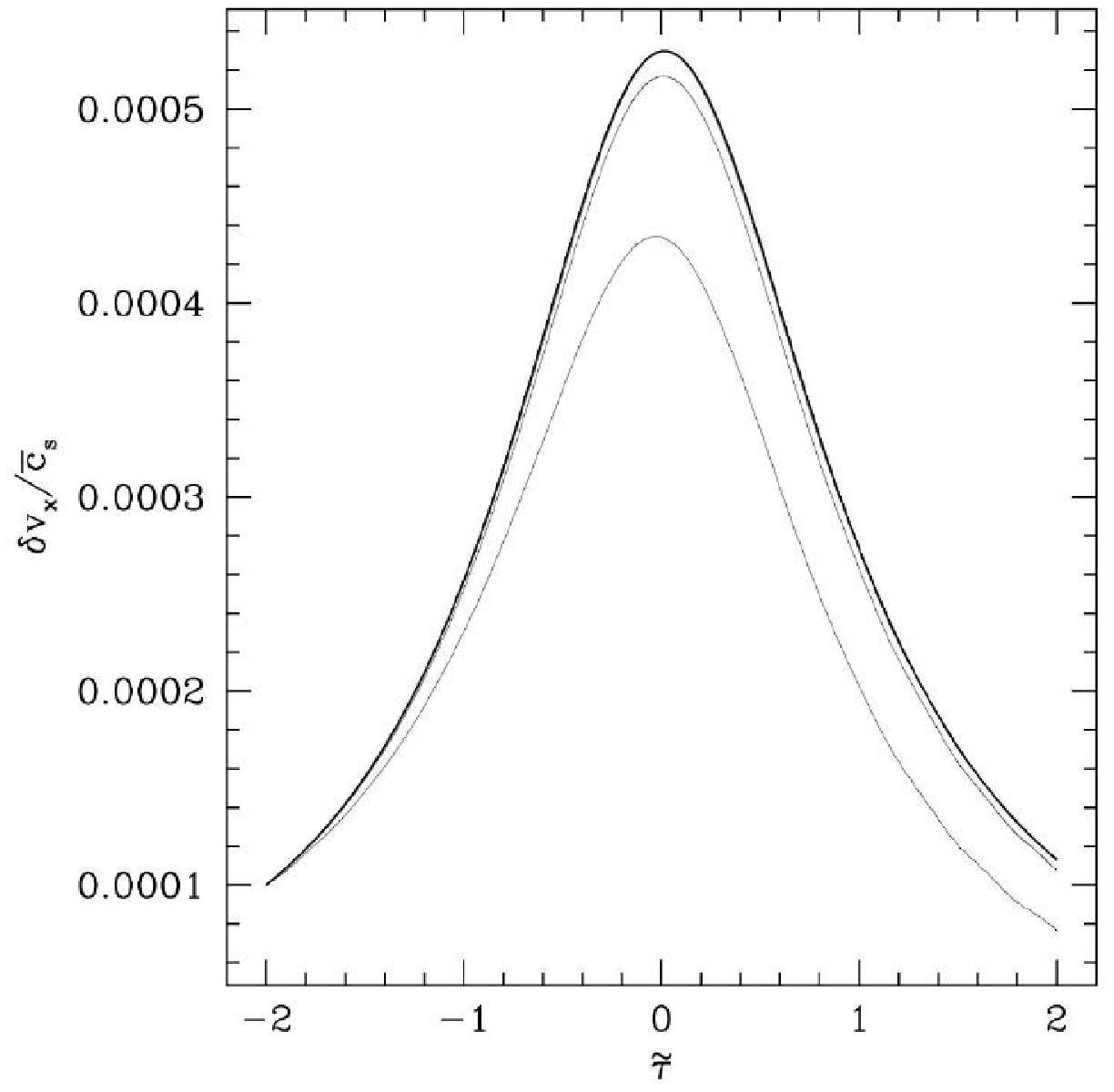}{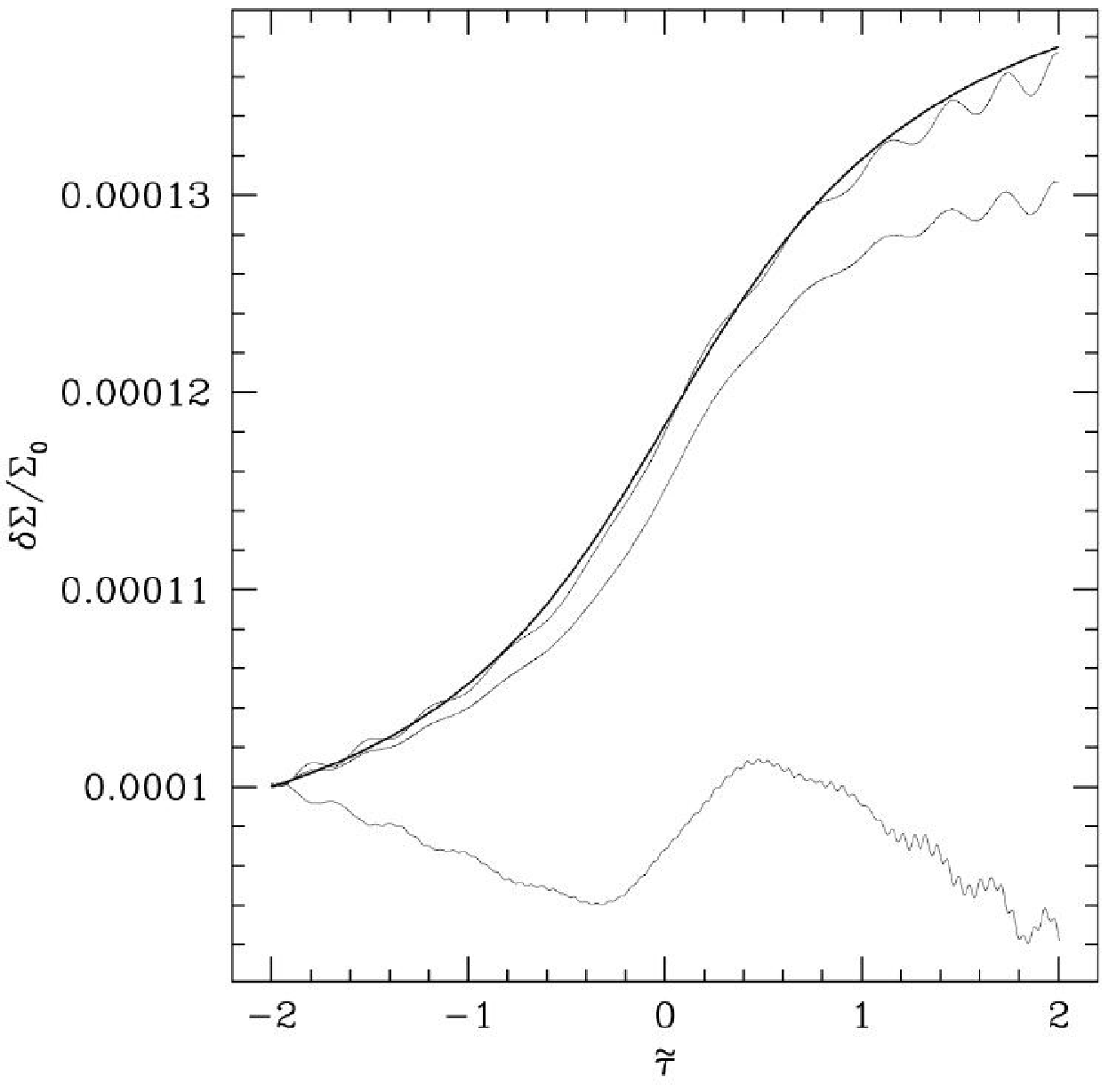}
\caption{
Evolution of a vortical shwave amplitude in the radially-stratified
shearing sheet (Run 1).  The left panel shows the radial velocity perturbation
and the right panel shows the density perturbation.  The heavy line in
each panel is the analytic result, and the light lines are runs with a
numerical resolution of (in order of increasing accuracy) $N_x \times N_y
= 1024 \times 16$, $2048 \times 32$ and $4096 \times 64$.  The number of
grid cells are chosen so that the shwave initially has the same number
of grid cells per wavelength in both the $x$ and $y$ directions.  The
results are shown for a test point at the minimum in $N_x^2$.
}
\label{f1}
\end{figure}

As a test for our code, we evolve a particular solution for the
incompressive shwaves in the radially-stratified shearing sheet (equations
[\ref{BOUSSVX2D}] and [\ref{XIY}]-[\ref{SOLDH}]).  The initial conditions
are $\delta v_x/\bar{c}_s = \delta \Sigma/\Sigma_0 = 1 \times 10^{-4}$ and
$k_{x0} = -128\pi/L_x$.  We set $L_y = 0.375$ and $k_y = 2\pi/L_y$ in
order to operate in the short-wavelength regime, and the other model
parameters are $q = 1.5$ and $\gamma - \Gamma = -0.3102$.  The latter
value yields a minimum value for $N_x^2(x)$ of $-0.01$.  The results of
the linear theory test are shown in Figure~\ref{f1}.

\section{Nonlinear Results}

Table 1 gives a summary of the runs that we have performed.  A
detailed description of the setup and results for each is given in the
following subsections.  Our primary diagnostic is a measurement of growth
rates, and the probe that we use for these measurements is an average over
azimuth of the absolute value of $v_x = \delta v_x$ at the minimum in
$N_x^2$.  Measuring $v_x$ allows us to demonstrate the damping of the
initial radial oscillations, and the average over azimuth masks the
interactions between multiple Fourier components with different growth rates in our
measurements. We will reference this probe with the following definition:
\be
v_t \equiv \langle |v_x(x_{min},y)| \rangle,
\ee
where here angle brackets denote an average over $y$ and $x_{min}$ is the
$x$-value at which $N_x^2(x)$ is a minimum.

\subsection{External Potential in Non-Rotating Frame}

As a starting problem, we investigate a stratified flow with $\bv_0 = 0$.
Such a flow can be maintained in equilibrium by replacing the tidal force
in equation (\ref{EQ2}) with an external potential $\Phi = -h_0$.  This can
be done in either a rotating or non-rotating frame.  It is a particularly
simple way of validating our study of convective instability in the shearing
sheet.  The condition $\bv_0 = 0$ implies $\qe = 0$, and therefore equation
(\ref{DRQ0}) should apply in the WKB limit, with the expected growth rate
obtained by evaluating equation (\ref{N2}) locally.\footnote{The fastest growing
WKB modes will be the ones with a growth rate corresponding to the minimum
in $N_x^2$.}  We have performed a fiducial run with an imposed external
potential in a non-rotating frame ($\Omega = 0$ in equation [\ref{EQ2}]) to
compare with the outcome expected from the Schwarzschild stability criterion
implied by equation (\ref{DRQ0}).  We set $\gamma - \Gamma = -0.3102$,
corresponding to $N_{x,min}^2 = -0.01$, and $L_y = L_x$.  The expected growth
rate for this Schwarzschild-unstable equilibrium is $0.1$ (in units of the
average radial sound-crossing time).  The numerical resolution for the fiducial
run is $512 \times 512$, and all the variables are randomly perturbed at an
amplitude of $\delta \Sigma/\Sigma_0 = \delta P/P_0 = \delta \bv/c_s = 1.0 
\times 10^{-12}$.

\begin{figure}
\plotone{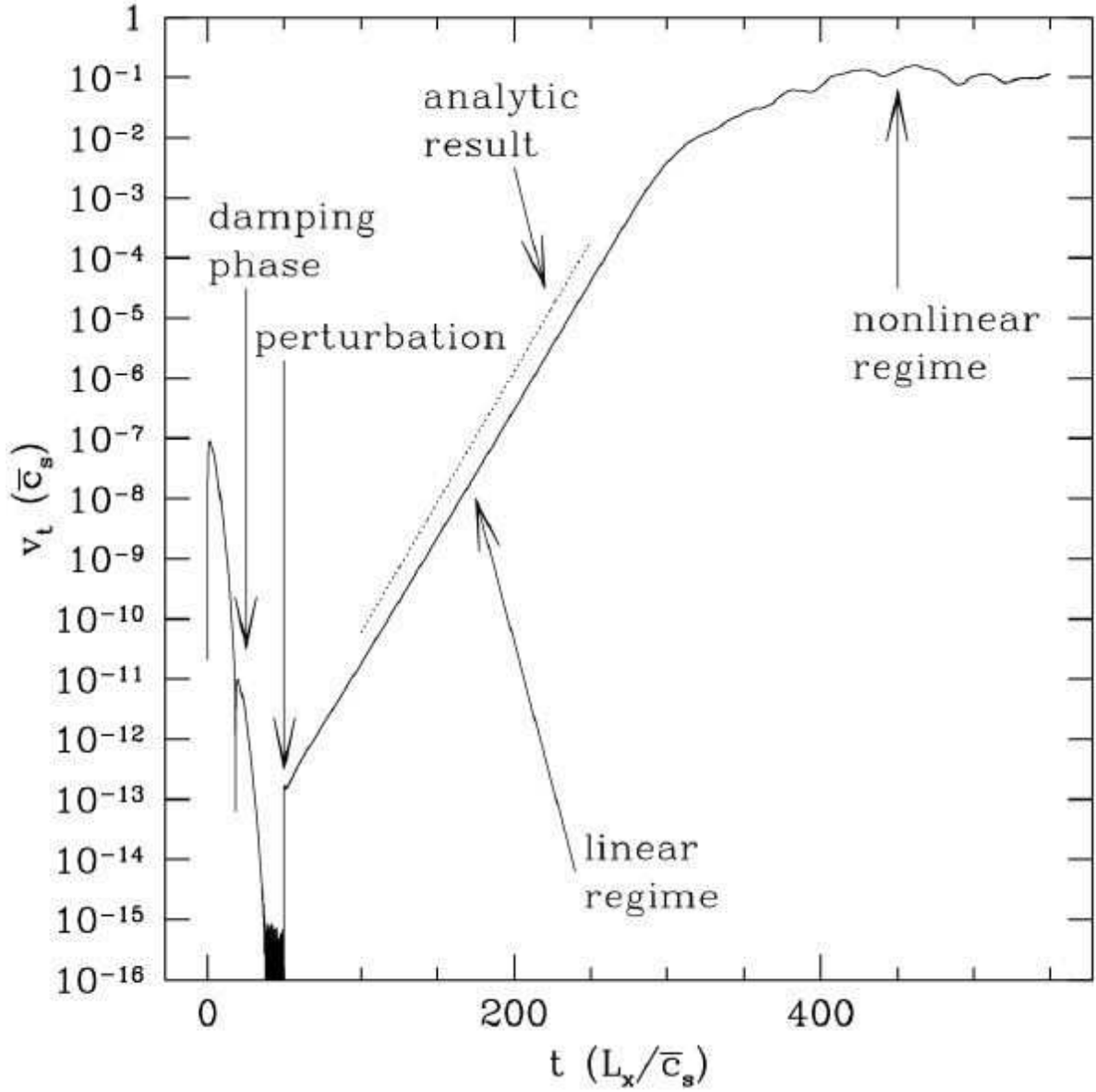}
\caption{
Evolution of $v_t$ as a function of time for Run 2 (external potential,
non-rotating frame).
}
\label{f2}
\end{figure}

\begin{figure}
\plottwo{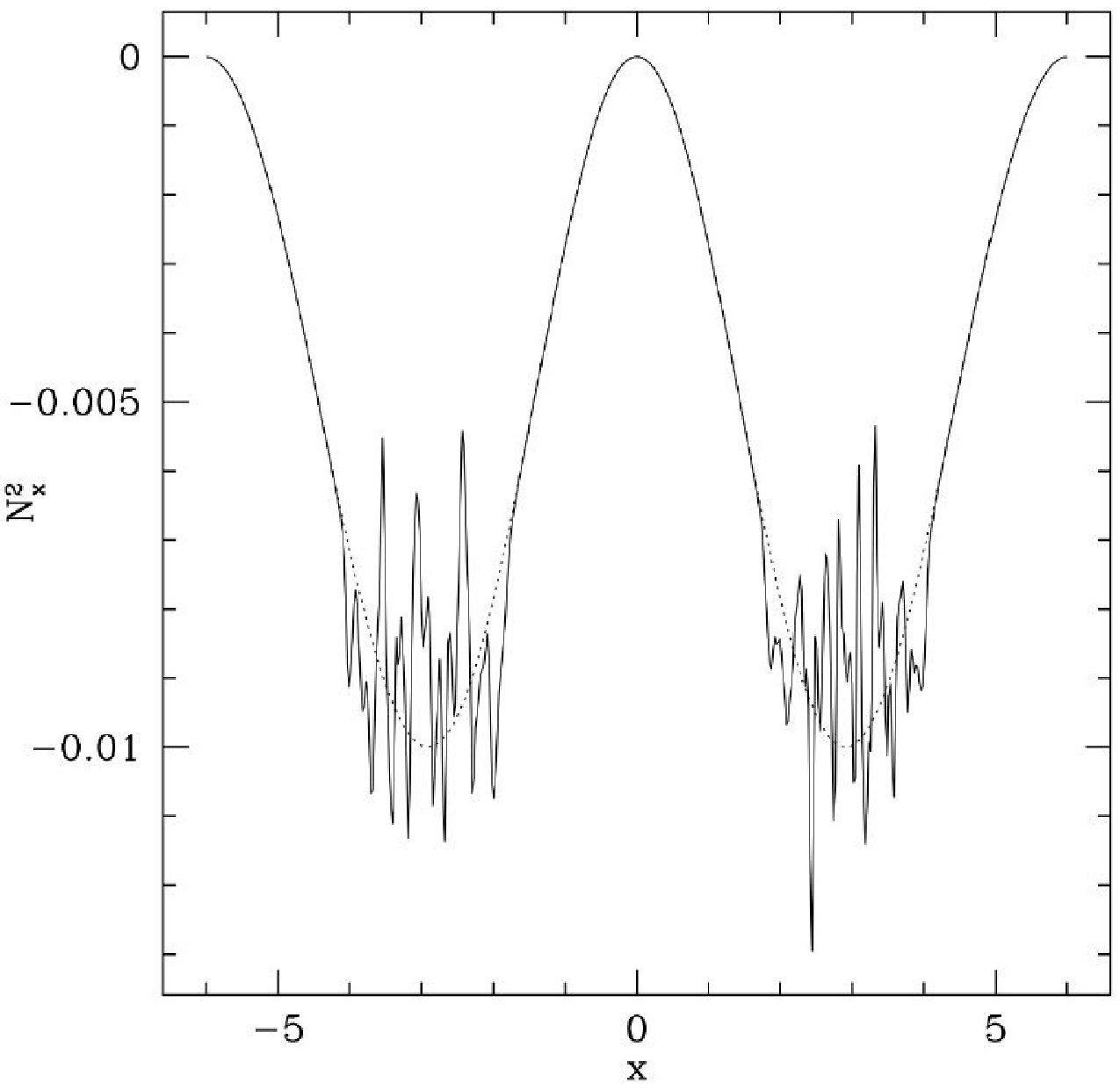}{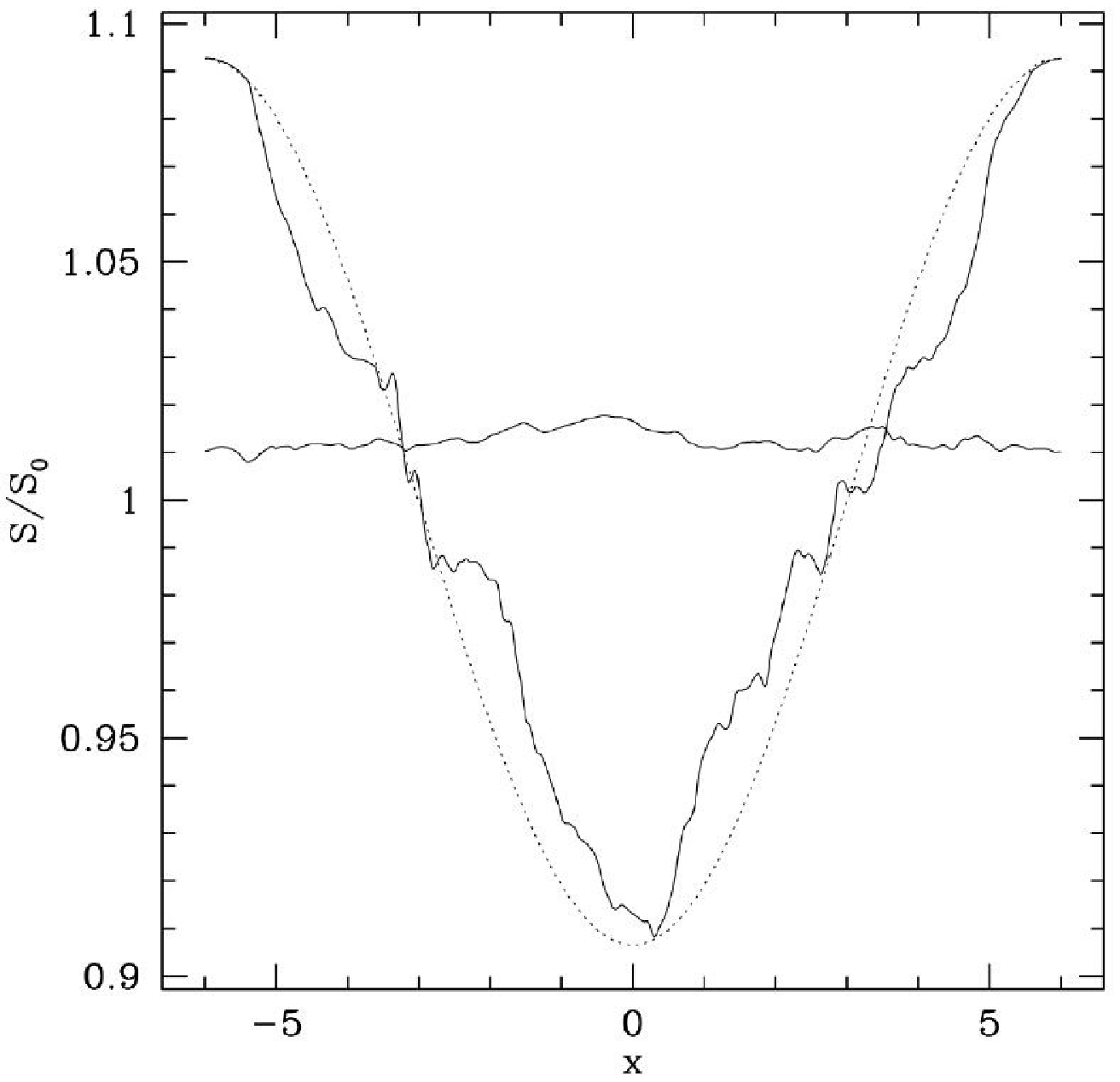}
\caption{
Plot of $N_x^2$ (left) and $S$ (right) as a function of $x$ for Run 2.  Both
are averaged over $y$.  The dotted line shows the equilibrium profile, and the
solid lines show snapshots during the nonlinear regime.  Growth initially occurs
at the minimum in $N_x^2$, and the entropy eventually settles to a nearly-constant
value.
}
\label{f3}
\end{figure}

\begin{figure}
\plotone{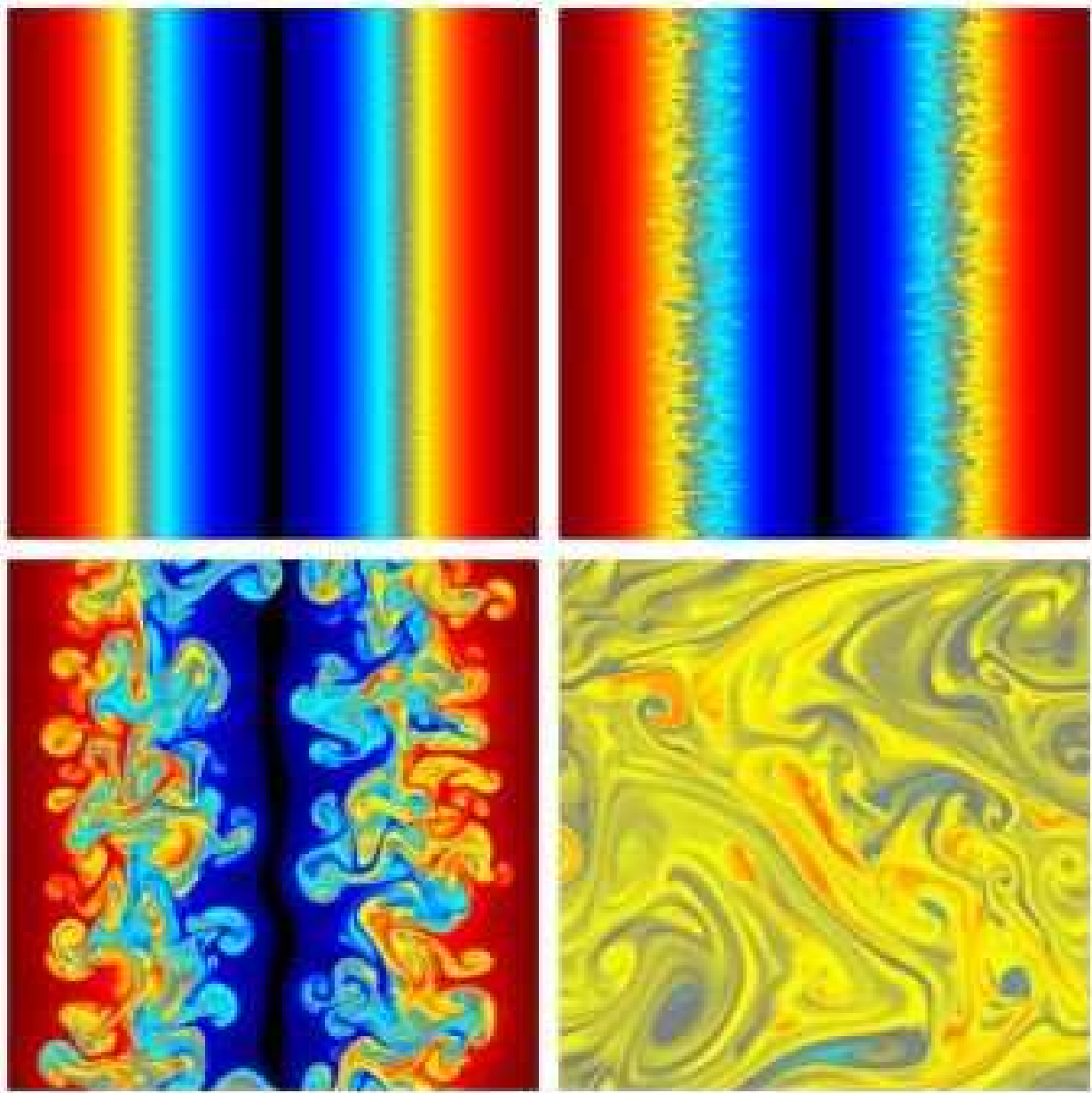}
\caption{
Snapshots of the entropy in the nonlinear regime for Run 2, indicating
maximum growth for modes near the grid scale and the eventual turnover of
the equilibrium entropy profile to its average value.  Dark shades indicate
values above (red in electronic edition) and below (blue in electronic edition) 
the average value.
}
\label{f4}
\end{figure}

A plot of $v_t$ as a function of time is given in Figure~\ref{f2}, showing
the initial damping followed by exponential growth in the linear regime. The
analytic growth rate is shown on the plot for comparison.  A least-squares
fit of the data in the range $100 \leq t\Omega \leq 250$ yields a measured growth
rate of $0.0978$.\footnote{Measurements of the growth rate earlier in the
linear regime or over a larger range of data yield results that differ from
this value by at most $5\%$.}  Figure~\ref{f3} shows a cross section of $N_x^2$
as a function of $x$ after the instability has begun to set in, along with cross
sections of the entropy early and late in the nonlinear regime.  The growth is
initially concentrated near the minimum points in $N_x^2(x)$.  Eventually the entropy
turns over completely and settles to a nearly constant value.   Figure~\ref{f4}
shows two-dimensional snapshots of the entropy in the nonlinear regime.  Runs with
the same equilibrium profile except $\gamma - \Gamma \geq 0$ are stable.  There
is also a long-wavelength axisymmetric instability that is present for
$\gamma - \Gamma < 0$ even in the absence of the small-scale nonaxisymmetric
modes.  We measure its growth rate to be $0.07$.  Due to the long-wavelength
nature of these modes, they are not treatable by a local linear analysis.

\subsection{External Potential in Rotating Frame}

\begin{figure}
\plotone{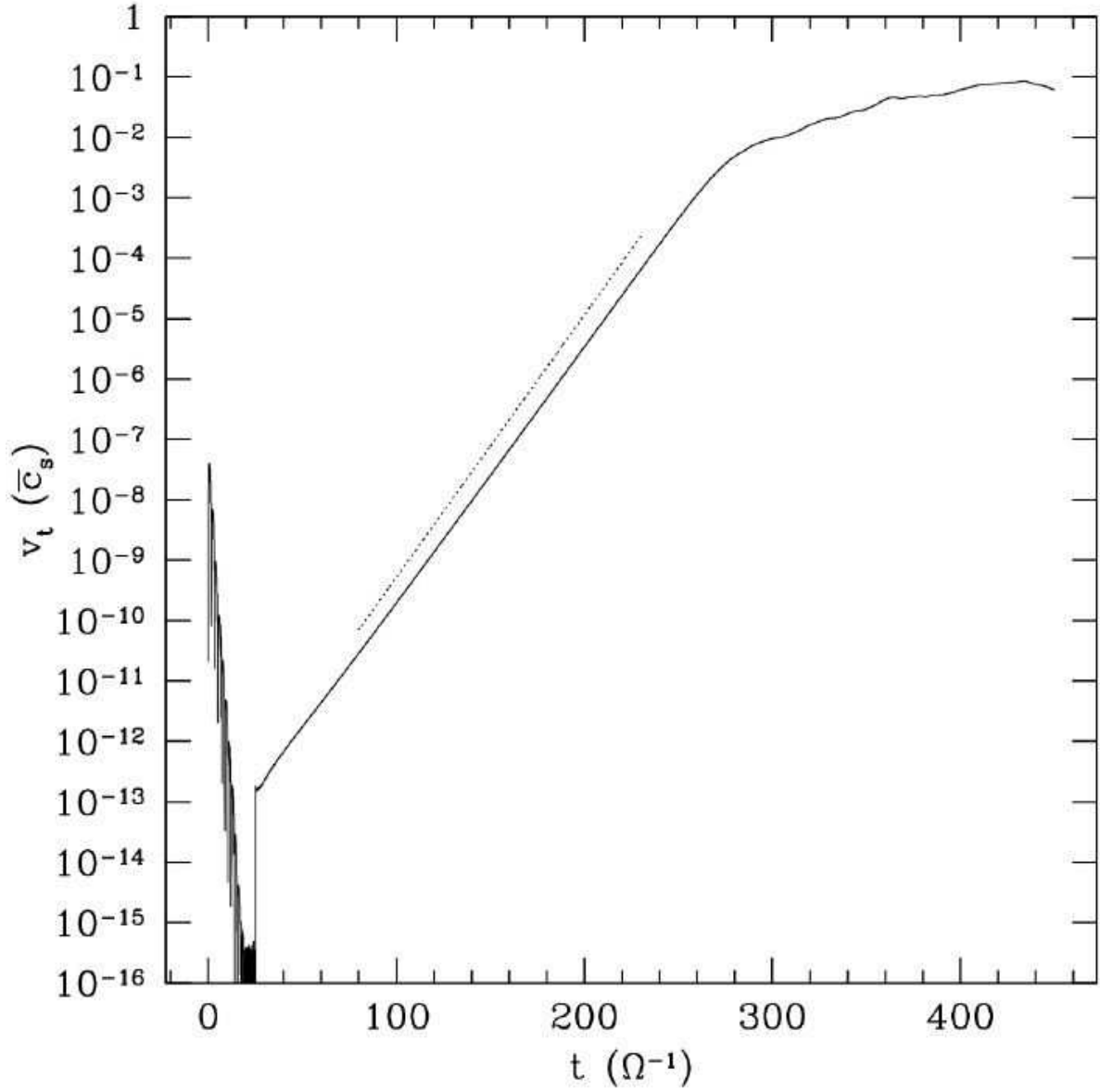}
\caption{
Evolution of $v_t$ as a function of time for Run 3 (external potential,
rotating frame).  The dotted line shows the expected growth rate.
}
\label{f5}
\end{figure}

\begin{figure}
\plotone{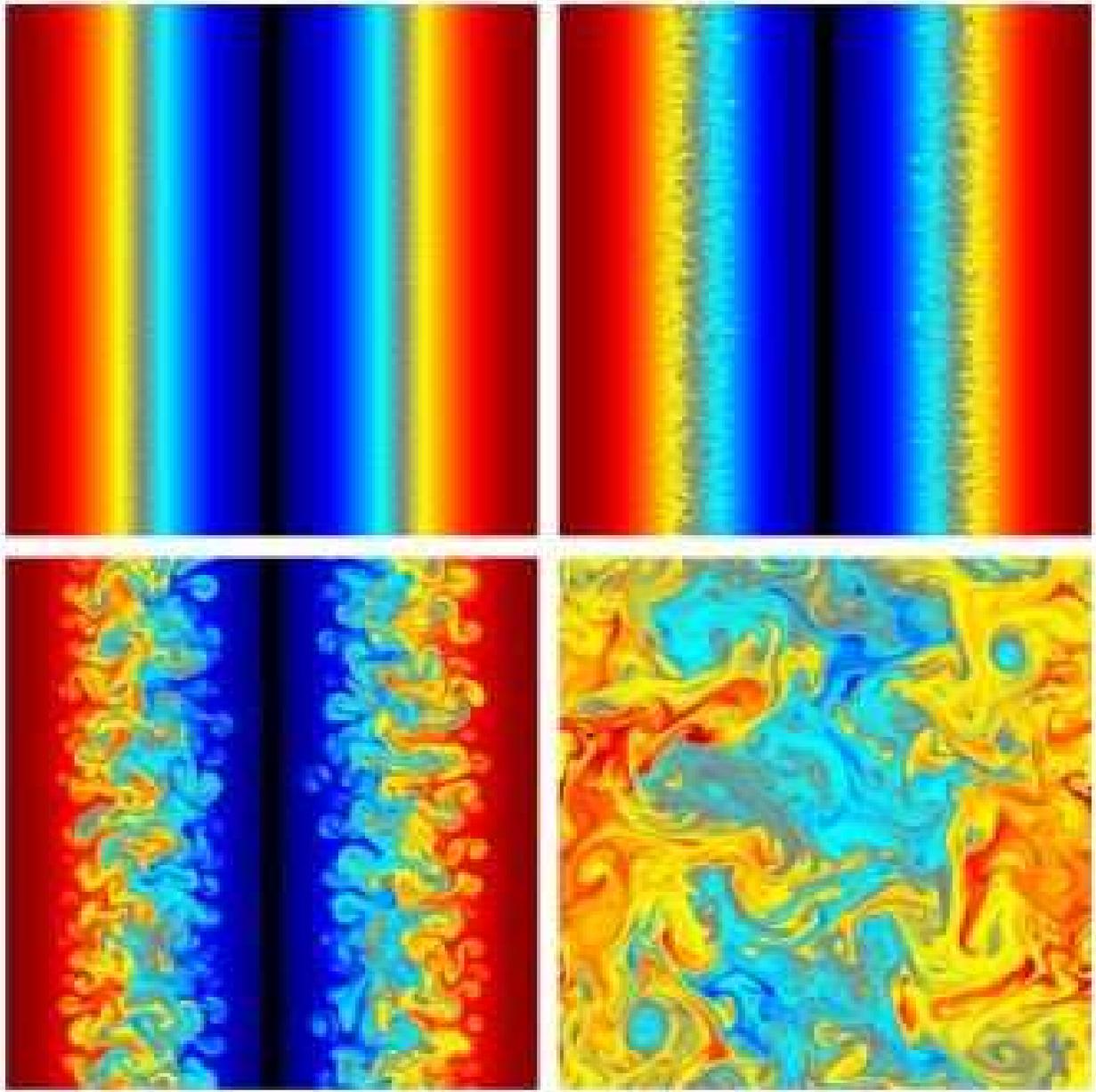}
\caption{
Snapshots of the entropy in the nonlinear regime for Run 3.  Dark shades indicate
values above (red in electronic edition) and below (blue in electronic edition) 
the average value.
}
\label{f6}
\end{figure}

We have performed the same test as described in \S5.1 in a rotating frame
($\Omega = 1$ in equation [\ref{EQ2}]).  Figure~\ref{f5} shows the
exponential growth in the linear regime for this run, with a measured growth
rate of $0.0977$.  Figure~\ref{f6} shows snapshots of the entropy in the
nonlinear regime.  The results are similar to the nonrotating case, except that
1) rotation suppresses the long-wavelength axisymmetric instability; 2) the
nonlinear outcome exhibits more coherent structures in the rotating case
including transient vortices; and 3) these coherent structures eventually
become unstable to a Kelvin-Helmholz-type instability.

\subsection{Uniform Rotation}

\begin{figure}
\plotone{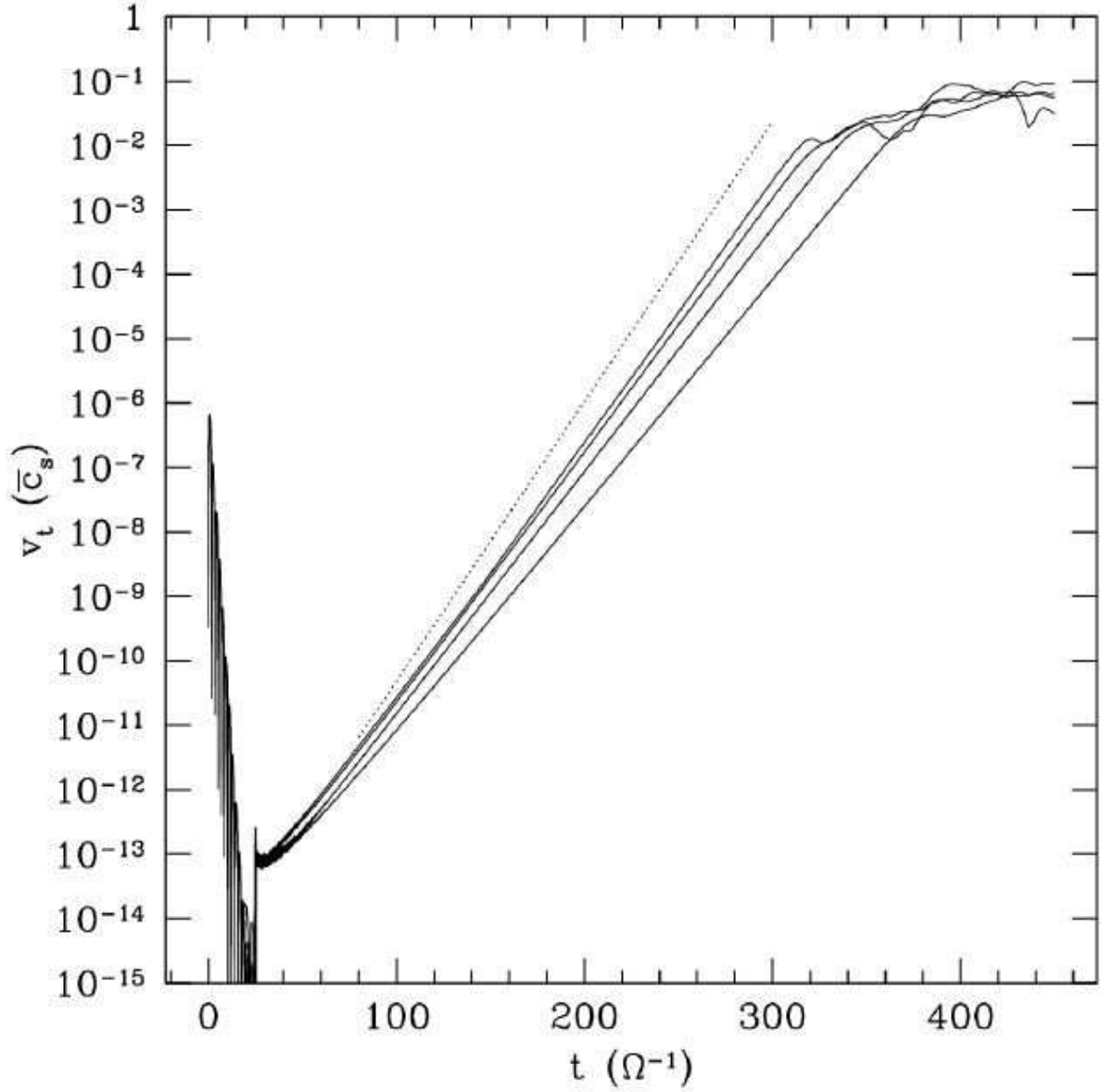}
\caption{
Evolution of $v_t$ as a function of time for Run 4 ($q = 0$).  The
dotted line shows the expected growth rate, and the solid lines are runs
with (in order of increasing growth) $L_y = 12$, $6$, $3$ and $1.5$.
}
\label{f7}
\end{figure}

\begin{figure}
\plotone{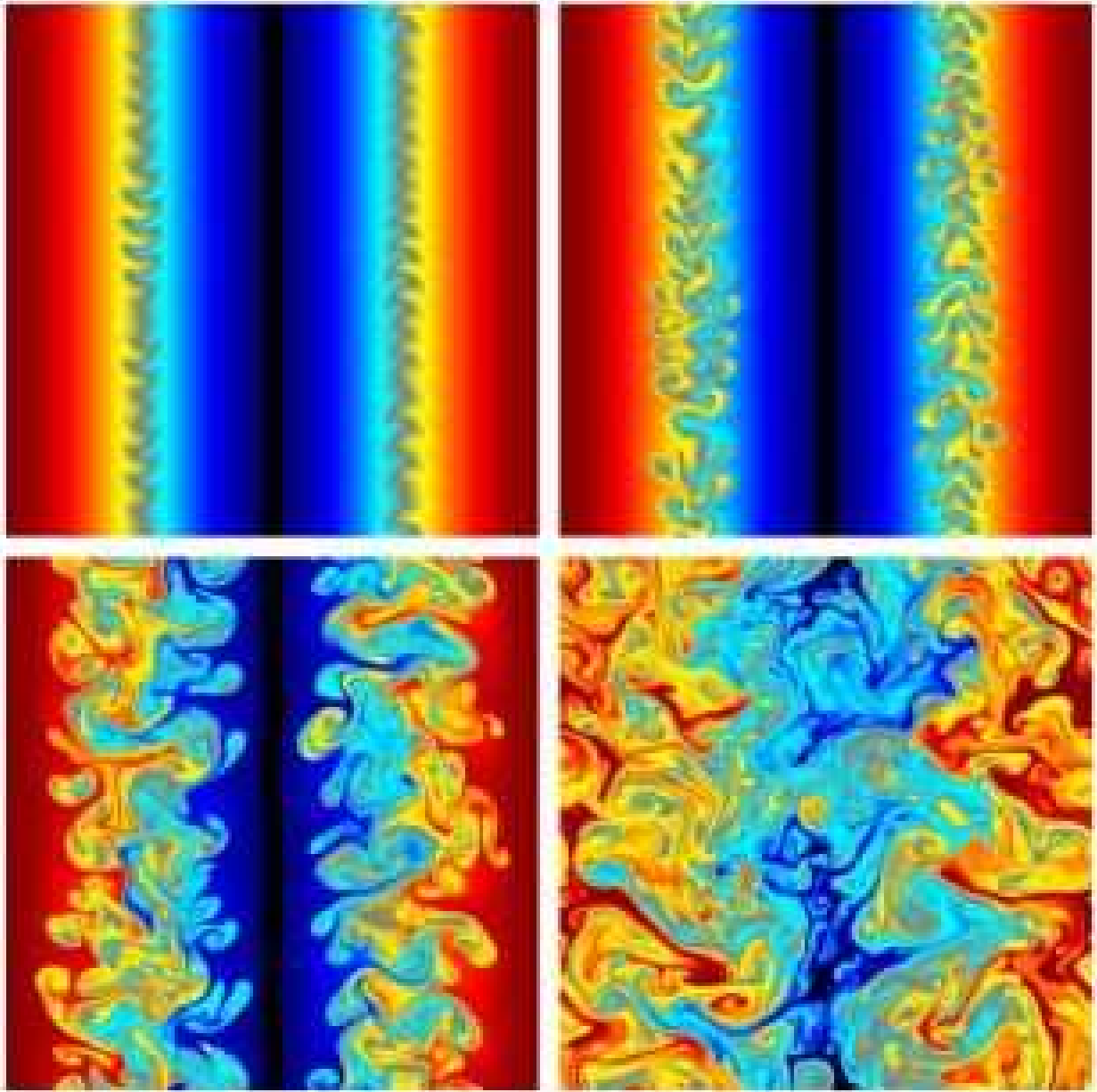}
\caption{
Snapshots of the entropy in the nonlinear regime for Run 4.  Dark shades indicate
values above (red in electronic edition) and below (blue in electronic edition) 
the average value.  Notice that the maximum growth does not occur for modes at 
the grid scale.
}
\label{f8}
\end{figure}

Having demonstrated the viability of simulating convective instability in
the local model, we now turn to the physically-realistic equilibrium
described in \S4.  We begin by setting the shear parameter $q$ to zero
in order to make contact with the results of \S\S5.1 and 5.2.  This is
analogous to a disk in uniform rotation.  The other model parameters are
the same as for the previous runs.  While there is still an effective
shear $-0.05 \lesssim \qe \lesssim 0.05$, near $\qe = 0$ one expects the
modes to obey equation (\ref{DRQ0}) in the WKB limit.  Figures~\ref{f7}
and \ref{f8} give the linear and nonlinear results for this run.  The
measured growth rate in the linear regime is $0.0809$.

\begin{figure}
\plotone{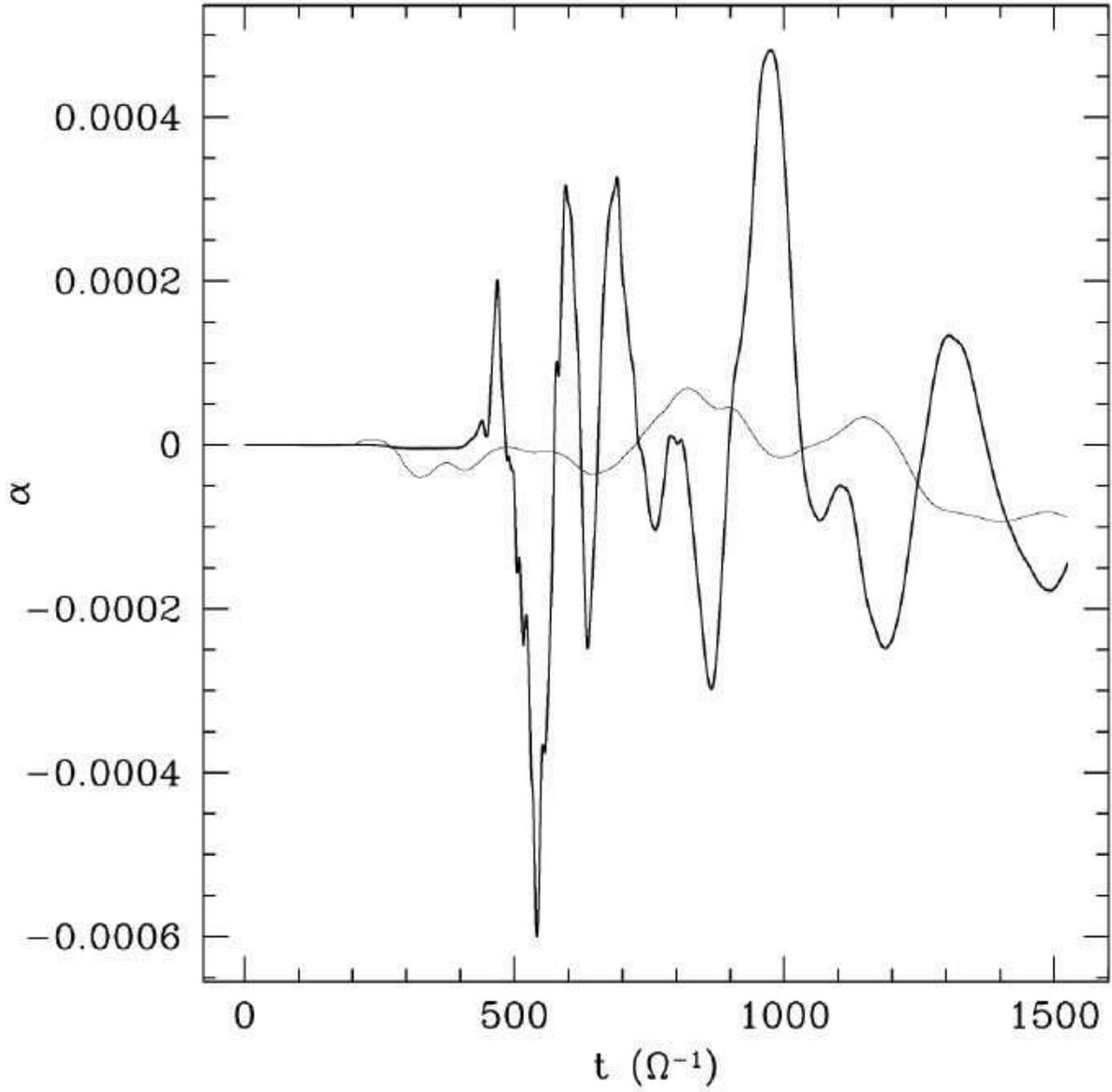}
\caption{
Evolution of the dimensionless angular momentum flux due to radial convection. 
The thin line is the data boxcar-smoothed over an interval $\Delta t = 500 
\Omega^{-1}$.
}
\label{f9}
\end{figure}

Consistent with results from numerical simulations of vertical convection
\citep{sb96,cab96}, the angular momentum transport associated with radial 
convection is slightly {\it inwards}.  Figure~\ref{f9} shows the evolution 
of the dimensionless angular momentum flux
\be
\alpha \equiv {1\over{L_x L_y \<P_0\>}} \int \Sigma \delta v_x \delta v_y dx dy,
\ee
where $\<P_0\>$ is the radial average of the equilibrium pressure, for an extended
version of Run 4.  The thin line is a version of the data that has been boxcar-smoothed 
to show the bias toward a negative flux.

There are two reasons for the larger error in the measured growth rate for
this run: 1) the equilibrium velocity gives rise to numerical diffusion due to
the motion of the fluid variables with respect to the grid; and 2) since the
growing modes are being advected in the azimuthal direction, the maximum growth
does not occur at the grid scale.  The latter effect can be seen in Figure~\ref{f8};
several grid cells are required for a well-resolved wavelength.  In order to resolve
smaller wavelengths, we have repeated this run with $L_y = 6$, $3$ and $1.5$.  The
results are plotted in Figure~\ref{f7} along with the results from the $L_y = 12$
run.  The measured growth rate for the $L_y = 1.5$ run is $0.0924$.

\begin{figure}
\plotone{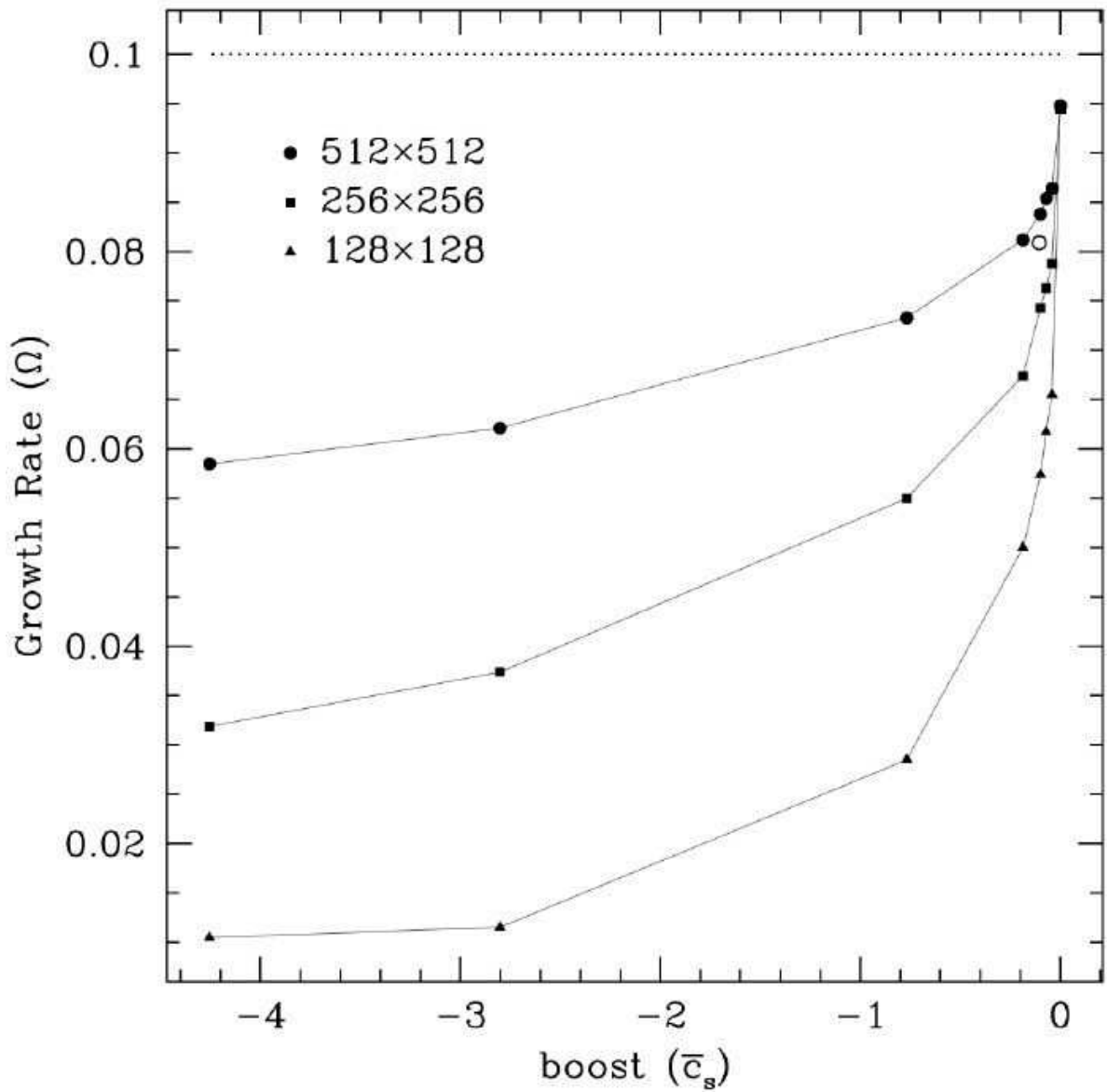}
\caption{
Growth rates as a function of azimuthal boost in a series of runs with
an external potential and $N_{x,min}^2 = -0.01$.  The dotted line shows
the analytic growth rate from linear theory.  The open circle denotes the
growth rate that was measured in Run 4, with the boost corresponding to
the magnitude of the velocity at the minimum in $N_x^2$ for Run 3 ($q = 0$).
The largest boost magnitude corresponds to the velocity at the minimum in
$N_x^2$ for Run 10 ($q = 1.5$).
}
\label{f10}
\end{figure}

To quantify the effects of numerical diffusion, we have performed a series
of tests similar to Run 2 (external potential in a rotating frame) but with
an overall boost in the azimuthal direction.  Figure~\ref{f10} shows measured growth
rates as a function of boost at three different numerical resolutions.  The
largest boost magnitude in this plot corresponds to the velocity at the
minimum in $N_x^2$ for a run with $q = 1.5$.  This highlights the importance 
of the fluid velocity with respect to the grid in determining numerical 
damping in ZEUS.

\subsection{Shearing Sheet}

\begin{figure}
\plotone{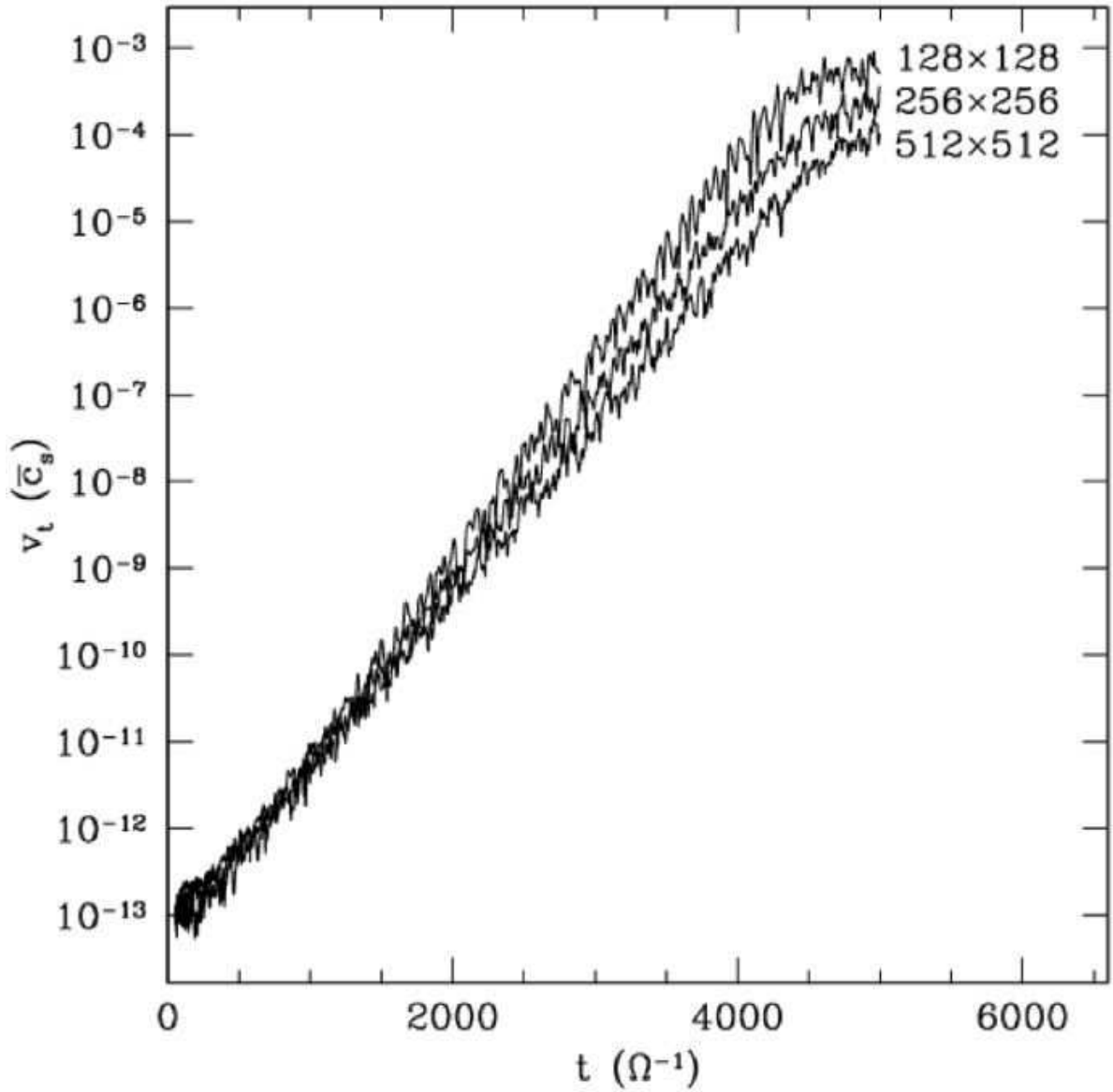}
\caption{
Evolution of $v_t$ as a function of time for Run 7 ($q = 0.2$ and $N_{x,min}^2 = -0.01$).
}
\label{f12}
\end{figure}

To investigate the effect of differential rotation upon the growth of this
instability, we have performed a series of simulations with nonzero $q$.
Intuitively, one expects the instability to be suppressed when the shear
rate is greater than the growth rate, i.e. for ${\rm Ri} \gtrsim -1$.
Figure~\ref{f11} shows growth rates from a series of runs with $N_{x,min}^2
= -0.01$ and small, nonzero values of $q$ at three numerical resolutions.  This
figure clearly demonstrates our main result: convective instability is suppressed by
differential rotation.  The expected growth rate from linear theory ($\sqrt{|N_x^2|}$
at $\qe = 0$) is shown in Figure~\ref{f11} as a dotted line.  If there is a
radial position where $\qe(x) = 0$ (i.e., ${\rm Ri} = -\infty$), $v_t$ at that
position looks similar to that of the previous runs (very little deviation from a
straight line); these measurements are indicated on the plot with solid points.
For $q \gtrsim 0.055$ there is no longer any point where $\qe(x) = 0$; in that case
$v_t$ was measured at the radial average between the minimum in $N_x^2(x)$ and the
minimum in $\qe(x)$, since this is where the maximum growth occured.  The data for
these measurements, which are indicated in Figure~\ref{f11} with open points, is
not as clean as it is for the runs with ${\rm Ri} = -\infty$ (see Figure~\ref{f12}).
All of the growth rate measurements in Figure~\ref{f11} were obtained by a least-squares
fit of the data in the range $1 \times 10^{-9} < v_t/\bar{c}_s < 1 \times 10^{-5}$.
The dashed line in Figure~\ref{f11} indicates the value of $q$ for which
${\rm Ri}_{min} = -1$.

Some of the growth in Figure~\ref{f11} appears to be due to aliasing.  This is a
numerical effect in finite-difference codes that results in an artificial transfer of
power from trailing shwaves into leading shwaves as the shwave is lost at the grid scale.
One expects aliasing to occur approximately at intervals of
\be\label{ALIAS}
\Delta \tilde{\tau} = \frac{N_x}{n_y}\frac{L_y}{L_x},
\ee
where $n_y$ is the azimuthal shwave number.  This interval corresponds to $\Delta
\tilde{k}_x(t) = 2\pi/dx$, where $dx = L_x/N_x$ is the radial grid scale.  Based upon
expression (\ref{ALIAS}), aliasing effects should be more pronounced at lower numerical
resolution because the code has less time to evolve a shwave before the wavelength of the
shwave becomes smaller than the grid scale.  It can be seen from the far-right data point
in Figure~\ref{f11} (Run 7 in Table~\ref{pap4t1}) that the measured growth rate
{\it decreases} with increasing resolution.  The evolution of $v_t$ for this run is
shown in Figure~\ref{f12}.

The effects of aliasing can be seen explicitly by evolving a single shwave, as was done for
our linear theory test (Figure~\ref{f1}).  Figure~\ref{f13} shows the evolution of
the density perturbation for a single shwave using the same parameters that were used for
Run 7: $L_x = L_y = 12$, $N_{x,min}^2 = -0.01$ and $q = 0.2$.
The initial shwave vector used was $(k_{x0}, k_y) = (-8\pi/L_x, 8\pi/L_y)$.  This
corresponds to $n_y = 4$, and the expected aliasing interval (\ref{ALIAS}) is therefore
$\Delta \tilde{\tau} = N_x/4$.  Runs at three numerical resolutions are plotted in
Figure~\ref{f13}, and the aliasing interval at each resolution is consistent with
expression (\ref{ALIAS}).  It is clear from Figure~\ref{f13} that a lower resolution
results in a larger overall growth at the end of the run.  It also appears that the growth
seen in Figure~\ref{f13} requires a negative entropy gradient.  We have performed this
same test with $N_x^2 > 0$, and while aliasing occurs at the same interval, there is no
overall growth in the perturbations.  This may be due to the fact that the perturbations
decay asymptotically for $N_x^2 > 0$ (see expression [\ref{ASYMP}]).

\begin{figure}
\plotone{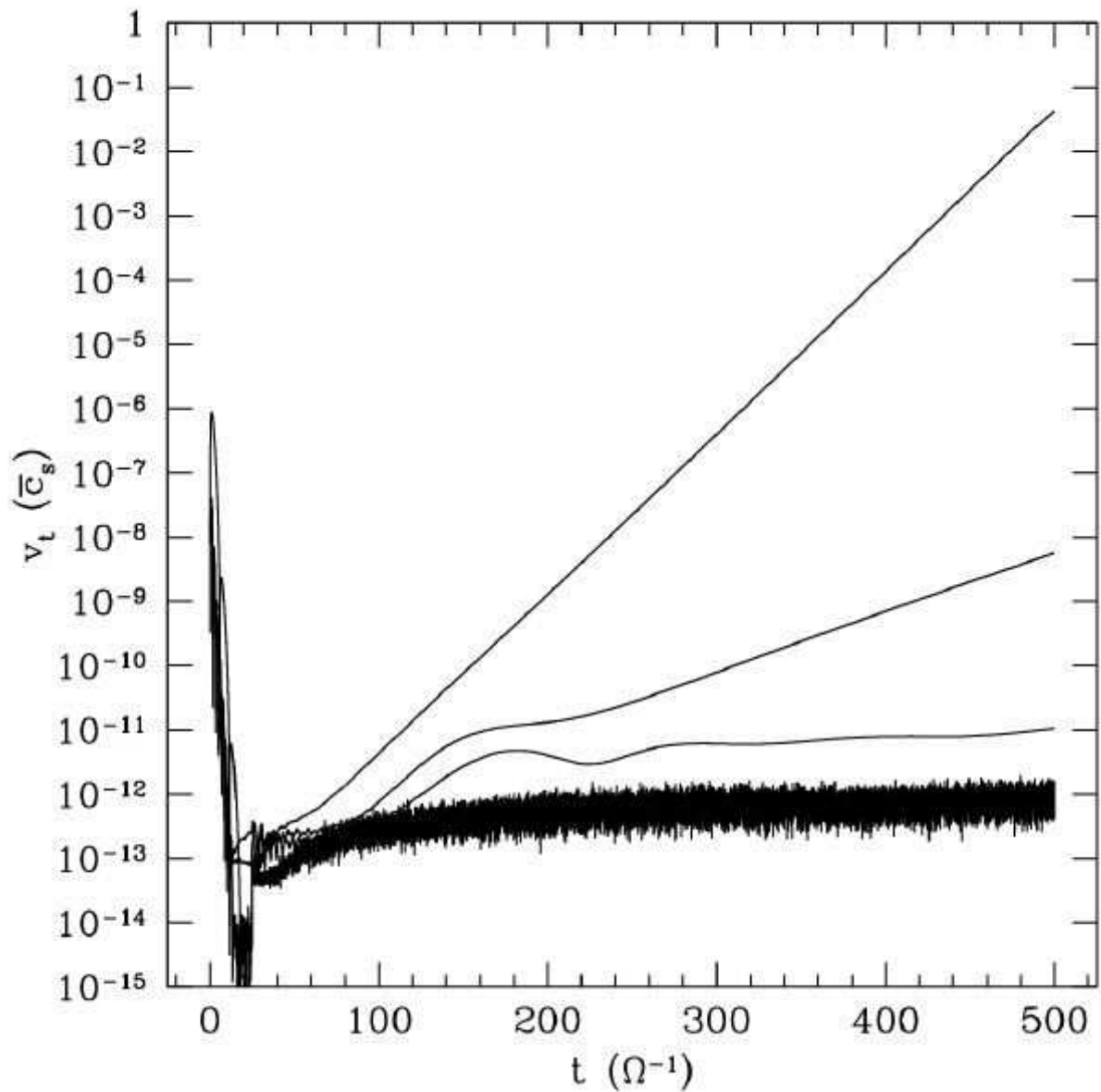}
\caption{
{Evolution of $v_t$ as a function of time for Run 10 ($q = 1.5$,
$N_{x,min}^2 = -0.01$).  Also shown are runs with $q = 0$ and an overall
boost equivalent to the velocity at the minimum in $N_x^2$ for Run 10, for
$N_{x,min}^2 = -0.01$ (measured growth rate of $0.058$), $N_{x,min}^2 = -0.003$
(measured growth rate of $0.021$) and $N_{x,min}^2 = -0.001$ (measured growth
rate of $0.0025$).}
}
\label{f15}
\end{figure}

Figure~\ref{f14} summarizes the parameter space we have surveyed, indicating that
there is instability only for $\qe \simeq 0$ and $N_x^2 < 0$.  The numerical
resolution in all of these runs is $512 \times 512$.  Figure~\ref{f15} shows the
evolution of the radial velocity in Run 10, a run with realistic
parameters for a disk with a nearly-Keplerian rotation profile and radial
gradients on the order of the disk radius: $q = 1.5$ and $N_{x,min}^2 =
-0.01$ (corresponding to ${\rm Ri} \simeq -0.004$).  Clearly no instability
is occurring on a dynamical timescale.  This plot is typical of all runs
for which the evolution was stable.  To give a sense for the minimum growth
rate that we are able to measure, we have also plotted in Figure~\ref{f15}
the results from several unstable runs with $q = 0$ and a boost equivalent to the
velocity at the minimum in $N_x^2$ for Run 10.  It is difficult to measure a growth rate for
the smallest value of $N_{x,min}^2$, but it is clear that there is activity present in this
run which does not occur in the stable run.  Based upon Figure~\ref{f15}, a
conservative estimate for the minimum growth rate that should be detectable in our
simulations is $0.0025 \Omega$.

\section{Implications}

\begin{figure}
\plotone{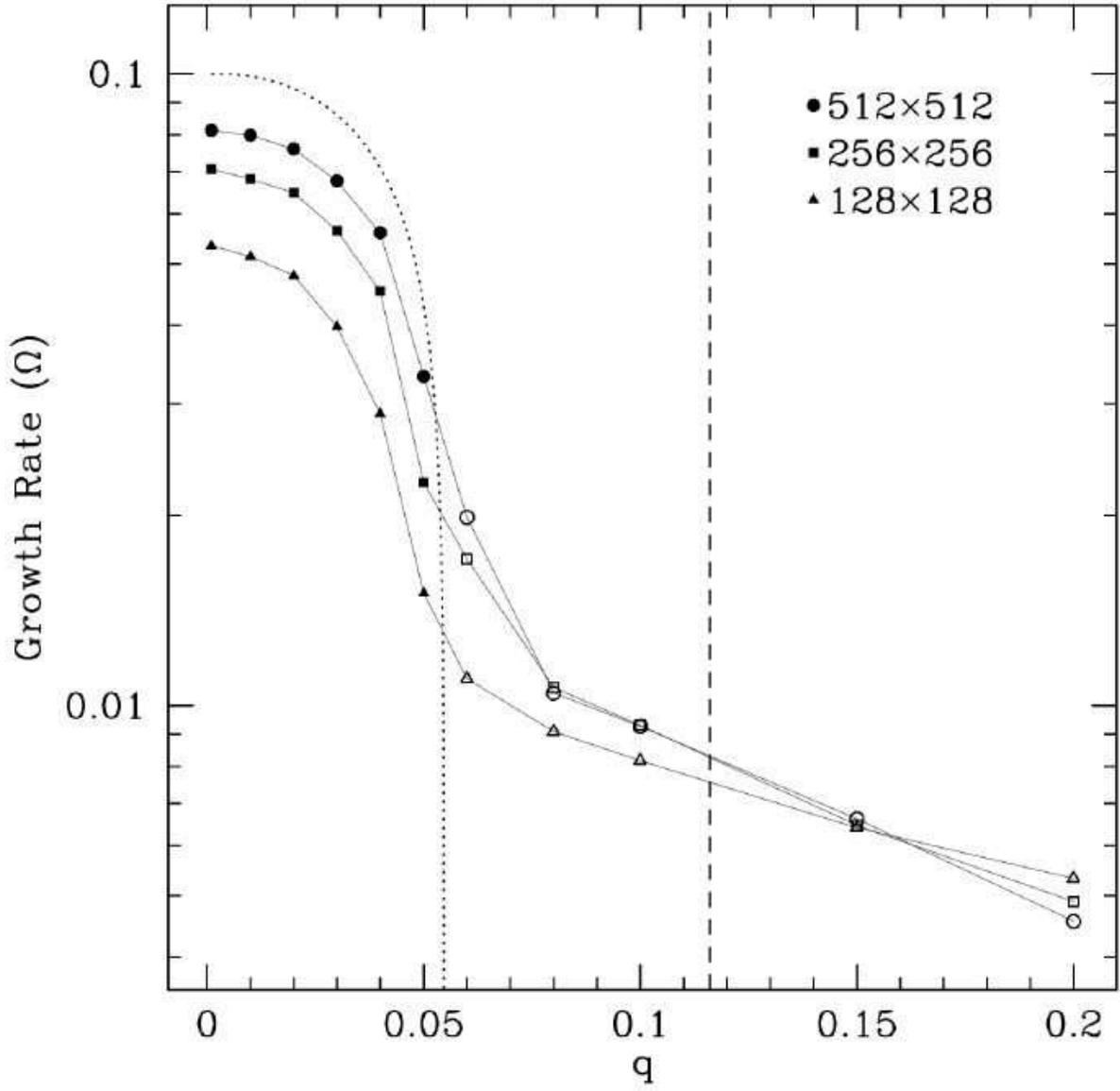}
\caption{
Growth rates as a function of $q$ with $N_{x,min}^2 = -0.01$.
See the text for a discussion.
}
\label{f11}
\end{figure}

\begin{figure}
\plotone{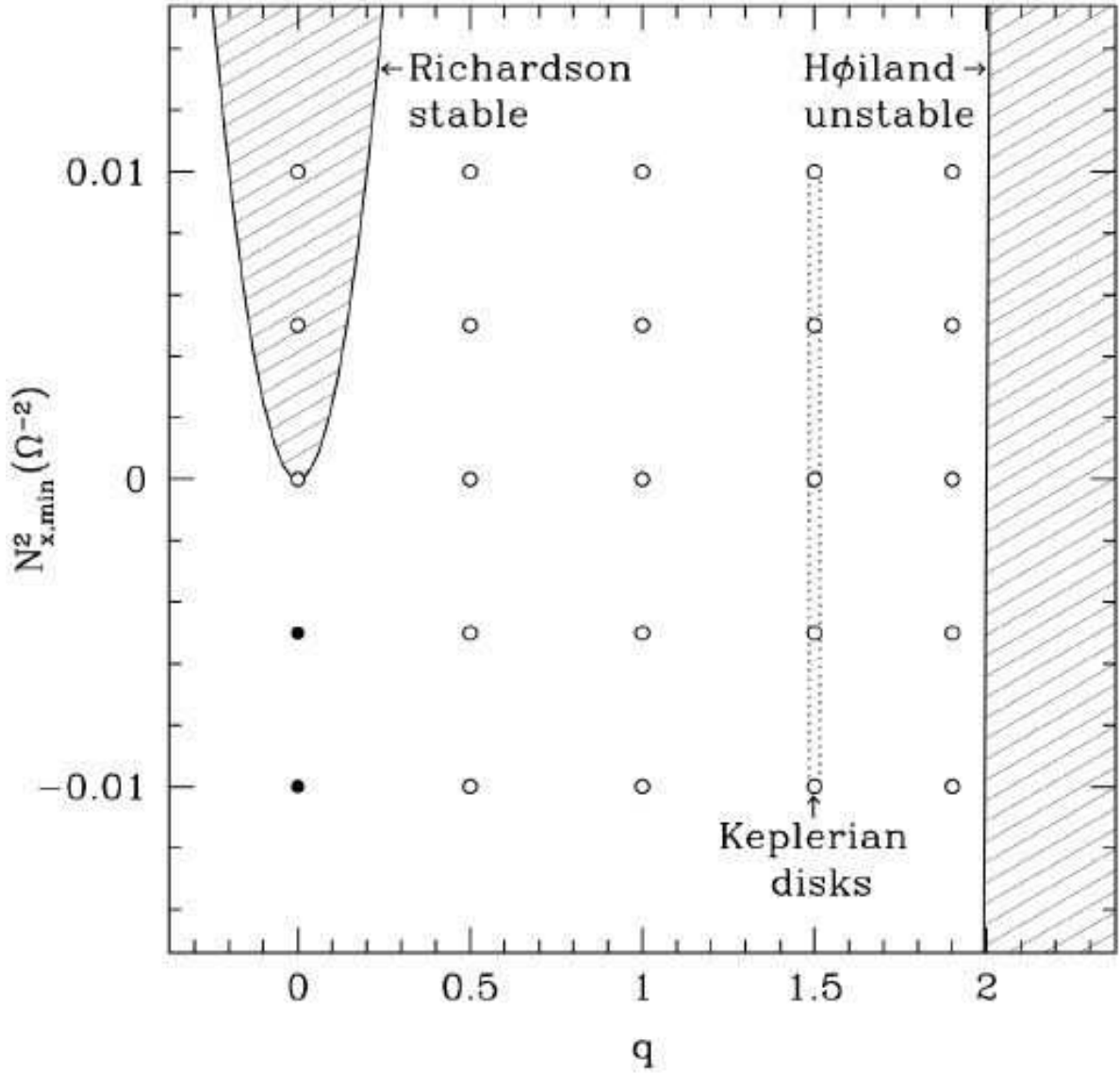}
\caption{
Parameter space surveyed in a search for nonlinear instabilities.  Closed
(open) circles denote runs that were unstable (stable).  The
only instability found was convective instability for $\qe \simeq 0$
and $N_x^2 < 0$ (${\rm Ri} \rightarrow -\infty$).  (We do not include on
this plot the runs shown in Figure~\ref{f11}.)
}
\label{f14}
\end{figure}

Our results are consistent with the idea that nearly-Keplerian disks are stable, 
or at most very weakly unstable, to local nonaxisymmetric disturbances.  Because 
our numerical resolution is finite, we cannot exclude the possibility of instability 
appearing at even higher resolution.  But Figure~\ref{f11}
demonstrates that convective instabilities that are present when the shear is nearly zero
are stabilized by differential rotation.  Perturbations simply do not have time to grow
before they are pulled apart by the shear.

\begin{figure}
\plotone{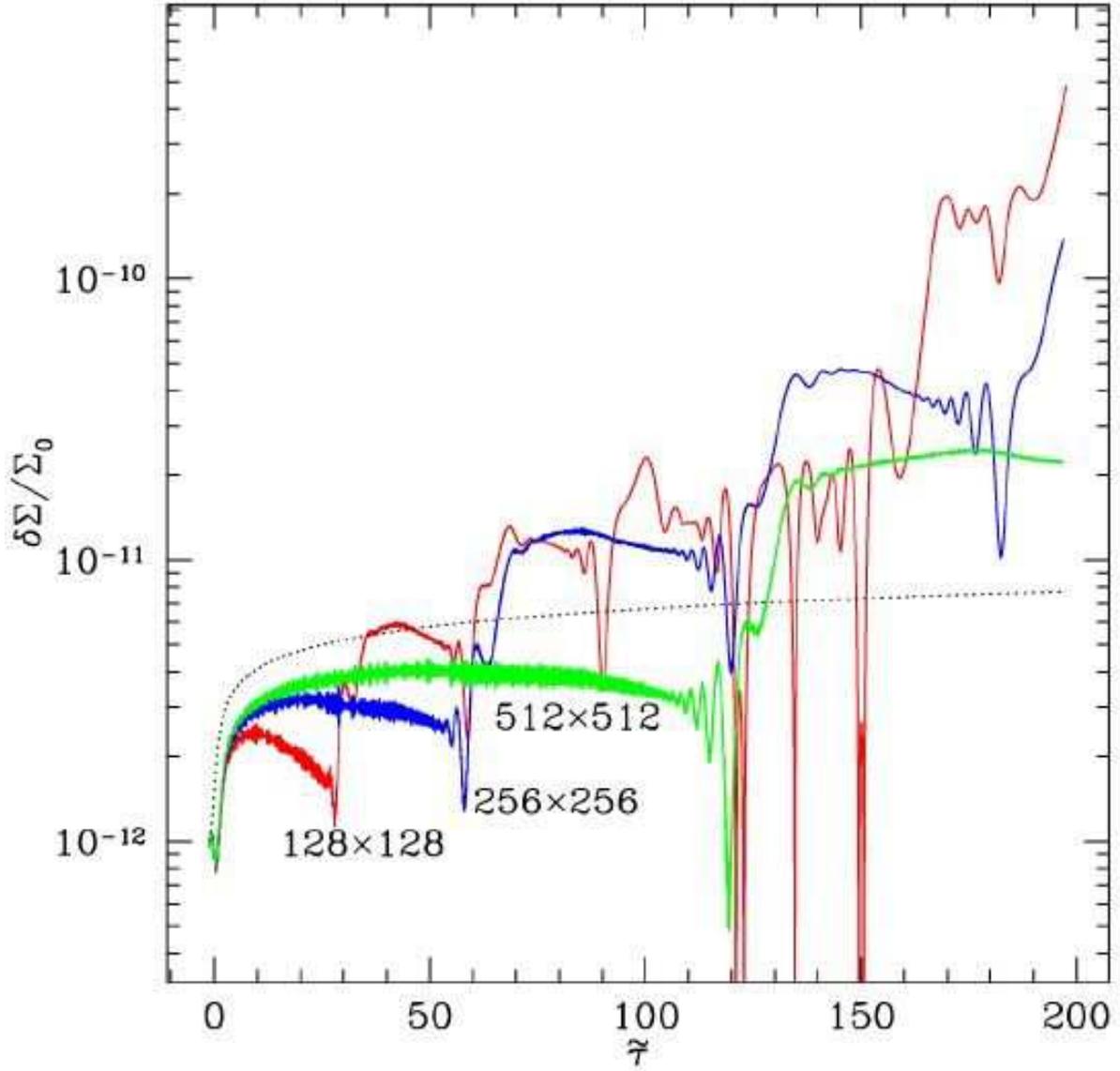}
\caption{
Evolution of the density perturbation for a single shwave with $q = 0.2$, 
$N_{x,min}^2 = -0.01$ and $L_y = L_x$ (Run 8).  The linear theory result 
is shown as a dotted line, along with results at three numerical resolutions: 
$128^2$, $256^2$ and $512^2$ (red, blue and green, respectively, in 
electronic edition).  Aliasing occurs when $\tilde{k}_x(t) = 2\pi/dx$.  
The overall growth, which is greater at lower numerical resolution, 
requires $N_x^2 < 0$.
}
\label{f13}
\end{figure}

An important implication of our results is that the instability claimed by \cite{kb03}
is {\it not} a linear or nonlinear local nonaxisymmetric instability.
Figure~\ref{f13} suggests that the results of \cite{kb03} may be due,
at least in part, to aliasing.  They use a finite difference code at fairly
low numerical resolution ($\leq 128^2$), and growth is only observed in
runs with a negative entropy gradient.  Curvature effects and the effects of
boundary conditions, which may also play a role in their global results,
cannot be tested in our local model.

In addition to being local, the simulations we have performed are two-dimensional. 
Interesting vertical structure is likely to develop in three-dimensional simulations, 
and this may affect our results.  At the same time, strong vertical stratification 
away from the midplane of the disk may enforce two-dimensional behaviour.

\cite{vmw98} calculated three-shwave interactions in the unstratified
(incompressible) shearing sheet and found that there is feedback from trailing
shwaves into leading shwaves for a small subset of initial shwave vectors.  It
would be of interest to revisit this calculation in the stratified shearing sheet.
One key difference between the stratified and unstratified shearing-sheet models
is that in the latter case all linear perturbations decay after their transient growth,
whereas in the former case the density perturbation does not decay.  Even though the
feedback in Figure~\ref{f13} is due to a numerical effect, these results show 
that feedback in the stratified shearing sheet can result in overall growth.

This work was supported by NSF grant AST 00-03091, PHY 02-05155,
NASA grant NAG 5-9180, and a Drickamer Fellowship for BMJ.

\newpage

\begin{deluxetable}{clcccc}
\tablenum{1}
\tablecolumns{6}
\tablewidth{0pc}
\tabcolsep 0.2truecm
\tablecaption{Summary of Code Runs \label{pap4t1}}
\tablehead{Run & Description & $L_y$ ($L_x = 12$) & $N_y$ & $N_x$ & Figure(s)}
\startdata
1 & Linear theory test & $0.375$ & $16, 32, 64$ & $64 N_y$ & \ref{f1} \\
2 & External potential, $\Omega = 0$ & $12$ & $512$ & $N_y$ & \ref{f2}-\ref{f4} \\
3 & External potential, $\Omega = 1$ & $12$ & $512$ & $N_y$ & \ref{f5}-\ref{f6} \\
4 & Uniform rotation ($q = 0$) & $12, 6, 3, 1.5$ & $512$ & $N_y$ & \ref{f7}-\ref{f9} \\
5 & External potential, $\Omega = 1$, boost & $12$ & $128, 256, 512$ & $N_y$ & \ref{f10} \\
6 & Small shear ($-\infty \lesssim {\rm Ri} \lesssim -1$) & $12$ & $128, 256, 512$ & $N_y$ & \ref{f11} \\
7 & Small shear ($q = 0.2, {\rm Ri} \gtrsim -1$) & $12$ & $128, 256, 512$ & $N_y$ & \ref{f12} \\
8 & Aliasing ($q = 0.2, {\rm Ri} \gtrsim -1$) & $12$ & $128, 256, 512$ & $N_y$ & \ref{f13} \\
9 & Parameter survey & $12$ & $512$ & $N_y$ & \ref{f14} \\
10 & Keplerian disk ($q = 1.5, {\rm Ri} \simeq -0.004$) & $12$ & $512$ & $N_y$ & \ref{f15} \\
\enddata
\end{deluxetable}

\newpage

\end{document}